\documentclass[conference]{IEEEtran}
\IEEEoverridecommandlockouts

\usepackage{cite}
\usepackage{amsmath,amssymb,amsfonts}
\usepackage{algorithmic}
\usepackage{graphicx}
\usepackage{textcomp}
\usepackage[dvipsnames]{xcolor}
\usepackage{xargs} 
\usepackage[utf8]{inputenc}
\usepackage{url}
\usepackage{cooltooltips}
\usepackage[colorinlistoftodos,prependcaption]{todonotes}
\usepackage{tikz}
\usepackage{gensymb}
\usepackage{fontawesome5}
\usepackage{xspace}

\usepackage[inline, shortlabels]{enumitem}
\usepackage{tabularx}
\usepackage{booktabs}
\usepackage{pgfplots}
\usepackage{wrapfig}
\usepackage{balance}
\usepackage{caption}
\usepackage{color,soul}
\usepackage[labelformat=simple]{subcaption}

\newcommandx{\unsure}[2][1=]{\todo[linecolor=red,backgroundcolor=red!25,bordercolor=red,#1]{#2}}
\newcommandx{\change}[2][1=]{\todo[linecolor=blue,backgroundcolor=blue!25,bordercolor=blue,#1]{#2}}
\newcommandx{\info}[2][1=]{\todo[linecolor=OliveGreen,backgroundcolor=OliveGreen!25,bordercolor=OliveGreen,#1]{#2}}
\newcommandx{\improvement}[2][1=]{\todo[linecolor=Plum,backgroundcolor=Plum!25,bordercolor=Plum,#1]{#2}}
\newcommandx{\thiswillnotshow}[2][1=]{\todo[disable,#1]{#2}}

\usepackage{tikz}
\usepackage{tkz-euclide}
  \usetikzlibrary{shapes,arrows,decorations.pathreplacing,calc,arrows.meta,automata,angles,shapes,decorations,patterns,positioning,intersections,backgrounds,fit}
\usepgfplotslibrary{fillbetween}

\definecolor{grass}{HTML}{296402}
\definecolor{way}{HTML}{373737}

\definecolor{pt_blue}{RGB}{0,119,187}
\definecolor{pt_cyan}{RGB}{51,187,238}
\definecolor{pt_teal}{RGB}{0,153,136}
\definecolor{pt_orange}{RGB}{238,119,51}
\definecolor{pt_red}{RGB}{204,51,17}
\definecolor{pt_magenta}{RGB}{238,51,119}
\definecolor{pt_grey}{RGB}{187,187,187}

\usepackage{tabularx}
\usepackage{multirow}
\usepackage{datatool}
\usepackage{pgfplots}
\usepackage{pgfplotstable}
\usepackage{etoolbox}
\usepackage{filecontents}
\usepackage{comment}
\usepackage[acronym]{glossaries}

\usepackage{cleveref} 
  \crefname{section}{Sect.}{Sects.}
  \Crefname{section}{Section}{Sections}
  \crefname{figure}{Fig.}{Figs.}
  \Crefname{figure}{Figure}{Figures} 
  \crefname{definition}{Definition}{Definitions}
  \crefname{equation}{}{}
  \Crefname{equation}{Equation}{Equations}
  \crefname{table}{Tab.}{Tabs.}
  \Crefname{table}{Table}{Tables}

\newcommand {\ie} {i.\,e.}
\newcommand {\eg} {e.\,g.}

\newcommand{\kw}[1] {\textsf{\texttt{#1}}}
\newcommand{\cbd} {CbD\xspace}

\newacronym{rl}{RL}{Reinforcement Learning}
\newacronym{dl}{DL}{Deep Learning}
\newacronym{cbd}{CbD}{Contract-based Design}
\newacronym{rm}{RM}{Runtime-Monitoring}
\newacronym{sos}{SoS}{systems of systems}
\newacronym{jml}{JML}{Java Modelling Language}
\newacronym{rqs}{RQs}{Research Questions}
\newacronym{rq}{RQ}{Research Question}
\newacronym{uml}{UML}{Unified Modelling Language}
\newacronym{lbc}{LbC}{Learning-based Components}
\newacronym{soa}{SOA}{Service-Oriented Architecture}
\newacronym{its}{ITS}{Intelligent Transportation Systems}

\tikzset{
    between/.style args={#1 and #2}{at = ($(#1)!0.5!(#2)$)}
}

\newcolumntype{P}[1]{>{\raggedright\arraybackslash}p{#1}}
\newcolumntype{R}[1]{>{\raggedright\arraybackslash}p{#1}}
\newcolumntype{L}[1]{>{\raggedleft\arraybackslash}p{#1}}
\newcolumntype{Y}{>{\centering\arraybackslash}X}

\newcommand{\error}[1][]{\ensuremath{\epsilon}\xspace}

\makeatletter
\newcommand\transformxdimension[1]{
    \pgfmathparse{((#1/\pgfplots@x@veclength)+\pgfplots@data@scale@trafo@SHIFT@x)/10^\pgfplots@data@scale@trafo@EXPONENT@x}
}
\newcommand\transformydimension[1]{
    \pgfmathparse{((#1/\pgfplots@y@veclength)+\pgfplots@data@scale@trafo@SHIFT@y)/10^\pgfplots@data@scale@trafo@EXPONENT@y}
}

\makeatother
\pgfplotsset{every tick label/.append style={font=\scriptsize}}

\def\BibTeX{{\rm B\kern-.05em{\sc i\kern-.025em b}\kern-.08em
    T\kern-.1667em\lower.7ex\hbox{E}\kern-.125emX}}
\begin{document}

\title{A Systematic Mapping Study on Contract-based Software Design for Dependable Systems
\thanks{This work has been funded by the Federal Ministry of Education and Research (BMBF) as part of MANNHEIM-AutoDevSafeOps (01IS22087P).}
}

\author{%
\IEEEauthorblockN{Fazli Faruk Okumus}
\IEEEauthorblockA{\textit{Research Institute AImotion Bavaria} \\
\textit{Technische Hochschule Ingolstadt}\\
Ingolstadt, Germany\\
fazlifaruk.okumus@thi.de}
\and
\IEEEauthorblockN{Amra Ramic}
\IEEEauthorblockA{\textit{Research Institute AImotion Bavaria} \\
\textit{Technische Hochschule Ingolstadt}\\
Ingolstadt, Germany}
\and
\IEEEauthorblockN{Stefan Kugele}
\IEEEauthorblockA{\textit{Research Institute AImotion Bavaria} \\
\textit{Technische Hochschule Ingolstadt}\\
Ingolstadt, Germany\\
stefan.kugele@thi.de}
}

\maketitle

\begin{abstract}
\emph{Background:}
\gls{cbd} is a valuable methodology for software design that allows annotation of code and architectural components with \emph{contracts}, thereby enhancing clarity and reliability in software development. It establishes rules that outline the behaviour of software components and their interfaces and interactions. This modular approach enables the design process to be segmented into smaller, independently developed, tested, and verified system components, ultimately leading to more robust and \emph{dependable software}.
\emph{Aim:}
Despite the significance and well-established theoretical background of \gls{cbd}, there is a need for a comprehensive systematic mapping study for reliable software systems. Our study provides an evidence-based overview of a method and demonstrates its practical feasibility.
\emph{Method:}
To conduct this study, we systematically searched three different databases using specially formulated queries, which initially yielded 1,221 primary studies. After voting, we focused on 288 primary studies for more detailed analysis. Finally, a collaborative review allowed us to gather relevant evidence and information to address our research questions.
\emph{Results:}
Our findings suggest potential avenues for future research trajectories in \gls{cbd}, emphasising its role in improving the dependability of software systems. We highlight maturity levels across different domains and identify areas that may benefit from further research.
\emph{Conclusion:}
Although \gls{cbd} is a well-established software design approach, a more comprehensive literature review is needed to clarify its theoretical state about dependable systems. Our study addresses this gap by providing a detailed overview of \gls{cbd} from various perspectives, identifying key gaps, and suggesting future research directions.
\end{abstract}
\section{Introduction}
\label{sec:introduction}
Given the dynamic nature of reliable systems, safety-critical applications continuously modify their software architecture during runtime.
This adaptability requires access to external services, including various system components and interfaces. However, Such access poses unique challenges related to reliability and trust, essential for ensuring safety and security. At this juncture, \gls{cbd} becomes notably significant.
\gls{cbd} offers a comprehensive way of creating this trust, especially if the contracts are explicitly formulated.

\gls{cbd} is an adaptable method that is applied at various stages of the design process.
It can be applied at the architectural level, where contracts are annotated to components, and at the code level, where contracts feature in code annotations.
The ability to bend to these varying requirements highlights the flexibility possessed by \gls{cbd}, a characteristic crucial in developing increasingly intricate \emph{dependable systems}.

The heterogeneous nature of these systems, which involves a complex software design, highlights the necessity for reliable design frameworks. Despite having a strong theoretical foundation and being readily available in academic literature, \gls{cbd} has not yet been widely adopted in industry \cite{tantivongsathaporn2006}.
There is a significant difference mainly due to the theoretical background of \gls{cbd}, which is based on Tony Hoare's early work. This includes Hoare logic~\cite{hoare:cacm:69} (\ie, a triple $\{P\}C\{Q\}$ of \emph{precondition} $P$, a command $C$, and postcondition $Q$) in 1969, and foundational ideas in program verification. These ideas culminate in the Eiffel programming language developed by Bertrand Meyer, incorporating many principles of \gls{cbd}, demonstrating its practical applicability \cite{mandrioli:meyer:aoose:91, meyer:computer:92}. Additionally, Benveniste~et~al.~\cite{benveniste:etal:fteda18} take a systemic approach.

This \emph{systematic mapping study} (SMS) aims to broaden the current body of knowledge in \gls{cbd} and assess its maturity, especially for dependable software systems by exploring its application across different levels of development, from architectural to code, and examining its integration in complex, heterogeneous system environments.
The detailed research questions in \cref{sec:study-design} are derived from the question:
\begin{quotation}
\textbf{What is the current body of knowledge in Contract-based Design?}    
\end{quotation}
This inquiry is complemented by examining the extent to which the methodology has been integrated across various application domains and the degree of maturity it has attained.
Building on this foundation, this study aims to identify and underscore emerging research directions that have not been adequately explored.
\paragraph{Contributions}
This research makes several scientific contributions, specifically:
\begin{enumerate}[(i)]
    \item The study provides an overview of the usage of \gls{cbd} theory within academia and dependable systems.
    \item It identifies gaps within the current body of knowledge.
    \item The research proposes future research directions, distinct from the identification of existing gaps.
    \item An assessment of the maturity level of \gls{cbd} theory across various domains is conducted, grounded in evidence from existing literature. 
\end{enumerate}
These points aim to deepen the understanding and guide the practical application of \gls{cbd}.
\paragraph{Outline}
The remainder of this paper is structured as follows: \Cref{sec:related-work} offers essential background information on contract-based design, setting the stage for a detailed discussion of related literature.
\Cref{sec:study-design} outlines the design of this study, including the posed research questions and the selection process.
The findings of the study are presented in \cref{sec:study-results}, where results are analysed based on gathered evidence.
We conclude with a comprehensive discussion and final observations in \cref{sec:discussion,sec:conclusion}, respectively.
\section{Background and Related Work}
\label{sec:related-work}
\subsection{Background}
\gls{cbd} represents a paradigm in software engineering aimed at constructing reliable software systems. Bertrand Meyer was instrumental in conceptualising this approach in 1992, concurrent with his development of the Eiffel programming language, as detailed in his seminal works \cite{mandrioli:meyer:aoose:91,meyer:computer:92,meyer:jss:88}.

In \gls{cbd}, software components are defined and interact based on specified \emph{contracts} within the overall system. These contracts typically involve class definitions, where methods may stipulate certain \emph{preconditions} (\ie, input assumptions) and \emph{postconditions} (\ie, guarantees about output). Additionally, software components uphold certain \emph{invariants} during runtime. This approach of \emph{Assume/Guarantee reasoning} frames each component's behaviour, interface, and responsibilities, fostering modularity and maintainability in software design.

The modularity of \gls{cbd} is a significant strength, as it segments the system into discrete, manageable subsystems governed by clearly defined interaction rules, thus defining its \emph{architecture}. This strength can be used primarily in \emph{\gls{soa}}, which is known for supporting the design of flexible systems and many works can be found that utilise \gls{soa} within the context of the automotive industry (cf.~\cite{DBLP:RodriguesPMACBV17,DBLP:KugeleOBCTH17}). This segmentation facilitates independent development and testing of each module, substantially easing the verification and testing process in complex systems, often referred to as \gls{sos}.

Moreover, \gls{cbd} enhances code readability by breaking the code into smaller, more manageable units. This decomposition not only aids in understanding the software but also serves as a form of documentation by explicitly defining component behaviours. Most mainstream programming languages lack native support for \gls{cbd}. However, libraries and plug-ins have been developed for multiple languages to integrate \gls{cbd} principles, such as the Java Modelling Language (JML) \cite{leavens1998jml}. Further details on using a programming language in \gls{cbd} will be discussed later in this study.
\subsection{Related Work}
While \gls{cbd} is a well-established software design approach, there needs to be more comprehensive literature synthesising its current state. Our study addresses this void by showing gaps in the literature and possible research direction. Also, it presents a detailed overview of \gls{cbd} theory from various perspectives. To the best of our knowledge, there has yet to be a systematic mapping study specifically focused on \gls{cbd} that comprehensively assesses the current landscape of this topic for dependable systems. Existing research predominantly revolves around various \gls{cbd} designs and implementations from a software engineering standpoint. This study, therefore, aims to provide a comprehensive overview of \gls{cbd} in the literature and highlight the potential gaps and research directions of this practice.
The following secondary studies are related to ours, examining the use of \gls{cbd} within the industry and quantitative studies demonstrating its impact on software design.

Tantivongsathaporn~et~al.~\cite{tantivongsathaporn2006} conducted a yearlong experiment on 16 software development teams that designed an industrial project within a corporation. The experiment compared \gls{cbd} practices with defensive programming practices.
Baudry~et~al.~\cite{baudry2001} conducted a study that demonstrates the positive effects of \gls{cbd} on designing \emph{object-oriented architectures}, particularly in terms of \emph{robustness} and \emph{diagnosability}.
In conclusion, the existing literature does not comprehensively depict the current state of \gls{cbd} theory. This study expands upon the existing body of work by providing a detailed examination of the current status of \gls{cbd} across various dimensions.

\section{Study Design}
\label{sec:study-design}
This section delineates the proposed research questions, their corresponding objectives, the criteria for selecting primary studies, methodologies for classifying these primary studies, and the specific search queries employed to source relevant primary studies from academic databases.

Our methodology is grounded in the well-established principles outlined in Kitchenham's guidelines \cite{Kitchenham}, which have been meticulously followed in conducting this study. In \cref{fig:selection}, our search and voting process implementation is depicted.

Before diving into the \gls{rqs}, this work aims to analyse the distribution of selected studies by years, citations, and venues, thereby gauging their impact and popularity in the field. Therefore, this work asks the following question to provide supplementary information:
\begin{enumerate}[left=0pt .. 2.3\parindent,label=\textbf{SI}]
\item\label{si:1}\textbf{What are the demographic details of the research in \cbd in terms of citations, publication year, venue?} 
\end{enumerate}
\subsection{Research Objectives and Questions}
The primary focus of this study is to examine the current landscape of contract-based design. This examination encompasses an analysis of design aspects and the maturity level of the existing primary studies. Moreover, this study extends its scope to evaluate how \gls{cbd} addresses solutions in areas like \gls{rl} and \gls{dl}. The guiding research questions for this study are as follows:

\SetLabelAlign{fixedwidth}{\hss\llap{\makebox[2.1em][l]{#1}}}
\begin{enumerate}[leftmargin=0pt, left = 0pt .. \parindent, itemindent=2em, label=\textbf{RQ\arabic*}, ]
    \item\label{rq:1} \textbf{What are the design details of \cbd?}
    Under this RQ, several sub-questions have been formulated to explore the nuances of contract design and its patterns:
    \begin{enumerate}[leftmargin=0pt, itemindent=2em, label=\textbf{RQ1.\arabic*}]
        \item\label{rq:1.1} \textbf{What varieties of \gls{cbd} are prevalent in the literature?} 
        This inquiry aims to shed light on the diverse types of \gls{cbd} employed in academic research.
        \item\label{rq:1.2} \textbf{Which programming languages are predominantly utilised for \cbd?}
        Recognising the limited native support for \gls{cbd} in programming languages, this question seeks to identify the tools and languages commonly used for implementing \gls{cbd}.
        \item\label{rq:1.3} \textbf{What proportion of the primary studies employs \gls{cbd}?}
        This question seeks to identify the industries where \gls{cbd} has been significantly applied.
    \end{enumerate}
    \item\label{rq:2} \textbf{What design solutions have been proposed for \gls{dl}, \gls{rl}, \gls{rm}, and mitigation strategies?}
    Given the stochastic nature of \gls{rl} and \gls{dl}, widely used in reliable systems, this research question focuses on their monitoring and management during operations, specifically the use of \gls{rm} and mitigation strategies for detected anomalies.
    \item\label{rq:3} \textbf{What is the maturity level in different areas?}
    This RQ is formulated to offer a comprehensive understanding of the primary studies and to categorise them in a way that elucidates their purpose and contribution. The maturity level assessment follows the guidelines and definitions introduced by Wieringa~et~al.~\cite{Wieringa}.
\end{enumerate}
\subsection{Search Strategy}
The study commenced with selecting digital databases to source primary studies, specifically ACM Digital Library (ACM DL), IEEE Xplore, and DBLP.
The search set of keywords and filters was determined through iterative refinement and leveraging a senior researcher's expertise in this field. Common keywords across databases included \kw{contract-based design}, \kw{design by contract}, and \kw{assume guarantee}. Additional keywords and filters were applied to refine the selection of relevant primary studies, as detailed in \cref{tab:search_strings}.
Each database required tailored approaches to optimise the relevance of the results. Keywords like \kw{Machine Learning}, \kw{Neural Network}, \kw{stochastic}, and \kw{probabilistic} were used to identify primary studies on systems with elements of uncertainty in both IEEE Xplore and ACM DL. Industry-specific keywords like \kw{robotics} and \kw{automotive} were employed to target primary studies in relevant application areas. The IEEE Xplore database also applied specific filters (see \cref{tab:search_strings}) to narrow the search results. Conversely, only general search queries were used for DBLP due to the lower paper yield.
\subsection{Selection of Relevant Primary Studies}
\begin{figure}
    \centering
 \scalebox{0.7}{   
    \begin{tikzpicture}[%
  node distance=1.0em and 0.5em,
  node/.style={draw, minimum height=2.3em, minimum width=3em, text width=3em, align=center, rounded corners = 2pt, font=\footnotesize\sffamily}     ,
  database/.style={cylinder, shape border rotate=90, draw, text width=4.5em, minimum height=2em, shape aspect=.25, align=center, font=\footnotesize\sffamily}
 ]
\node[node, text width=16.5em, text centered, fill=gray!30!white](ss) {Search strings};
%
\node [database, below =of ss.south west, anchor=north west] (acmdl) {ACM DL\\(353)};
\node (dblp) [database, below=of ss.south east, anchor=north east] {DBLP\\(399)};
\node (ieeexplore) [database, between=acmdl and dblp] {IEEE Xplore (469)};
\draw [<-](acmdl) -- (ss.south -| acmdl.north);
\draw [->](ss) -- (ieeexplore);
\draw [<-](dblp) -- (ss.south -| dblp.north);
%
\node[node, below left=of acmdl, fill=gray!10!white, text width=6em](dupes) {Duplicate removal};
\node[node, minimum width=1cm, below=of ieeexplore](dupesNr){1,221};
\draw [->](acmdl) |- (dupesNr);
\draw [->](ieeexplore) -- (dupesNr);
\draw [->](dblp) |- (dupesNr);
%
\node[node, minimum width=1cm, below=of dupesNr](title){1,077};
\node[node, below=of dupes, fill=gray!10!white, text width=6em](vtitle) {Voting based\\on title};
\draw [->](dupesNr) -- (title);
%
\node[node, minimum width=1cm, below=of title](abstract){474};
\node[node, below=of vtitle, fill=gray!10!white, text width=6em](vabstract) {Voting based\\on abstract};
\draw [->](title) -- (abstract);
%
\node[node, minimum width=1cm, below =of abstract](full){372};
\node[node, minimum width=1cm, below =of vabstract, text width=6em, fill=gray!10!white](vfull) {Voting based on full text};
\draw [->](abstract) -- (full);
\draw [->](vabstract) -- (vfull);
%
\node[node, minimum width=1cm, below =of full](exclusion){338};
\node[node, minimum width=1cm, below =of vfull, text width=6em, fill=gray!10!white](vexclusion) {Exclusion criteria};
\draw [->](full) -- (exclusion);
\draw [->](vfull) -- (vexclusion);
\node[node, minimum width=1cm, below =of exclusion](last){288};
\draw [->](exclusion) -- (last);
\node[node, minimum width=1cm, text width=3.7em,fill=gray!30!white](total) at (last -| dblp){Total};
\draw [->](last) -- (total);
\draw [->](dupes) -- (vtitle);
\draw [->](vtitle) -- (vabstract);
\end{tikzpicture}}
    \caption{Summary of the selection process.}
    \label{fig:selection}
\end{figure}
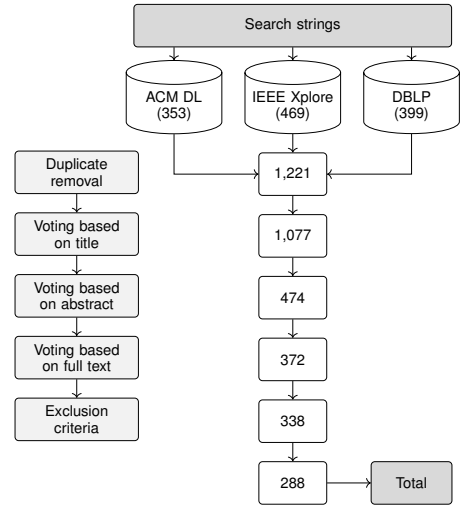

The initial search yielded 1,221 primary studies, which were collected on 10 July 2023. These studies were then reviewed for duplicates, and any identical primary studies were removed. Some results did not exactly match the search keywords used (see \cref{tab:search_strings}). To identify relevant primary studies, two of the authors independently reviewed titles, abstracts, and full texts, and reached a consensus through voting. Cohen's $\kappa$ \cite{mchugh2012interrater} was used to measure interrater reliability, with an initial $\kappa$ of 0.58 (moderate agreement) and a second stage $\kappa$ of 0.88 (almost perfect agreement). Finally, the authors conducted a final vote based on reviewing the full texts of the primary studies to reach a consensus. The entire process is summarised in \cref{fig:selection}.
After applying the exclusion criteria from \cref{tab:exclusion}, we narrowed it down to 288\footnote{In case of acceptance, the papers will be mentioned and references added.} primary studies for detailed analysis.
\begin{table}[!t]
    \centering
    \caption{Exclusion criteria}
    \label{tab:exclusion}
    \begin{tabularx}{\columnwidth}{@{}R{0.2cm}X@{}}
    \toprule
    \textbf{Nr.} & \textbf{Exclusion Criteria}\\
    \toprule
    1 & Papers that are not related to \gls{cbd} \\
    2 & Papers that are not written in English \\
    3 & Papers that are not available to download\\
    4 & Front Matter, Back Matter, Proceedings Descriptions, Full Proceedings\\
    5 & Preprints, if there is a published version in a journal or conference\\
    6 & Papers that are a duplicate \\
    7 & Papers that are related to economics, social sciences, games, blockchain, cloud computing, quantum computing, or medicine  \\
    8 & Teaching modules \\
    \bottomrule
    \end{tabularx}
\end{table}
\subsection{Data Extraction and Classification Scheme} 
We thoroughly examined each study's title, abstract, and conclusion to address the \gls{rqs} and extract pertinent information from the 288 primary studies. Additionally, we reviewed and selectively read the complete texts. A detailed account of the specific information sought in each study relative to every research question is provided in \cref{tab:dataext}.
\begin{table}[t!]
    \centering
        \caption{Data extraction details}
        \label{tab:dataext}
        \begin{tabularx}{\columnwidth}{@{}R{1cm}X@{}}
        \toprule
        \textbf{RQ}  & \textbf{Description} \\
        \midrule

        \textbf{\ref{rq:1.1}}     & Classify types of \gls{cbd} based on component descriptions. \\
        \textbf{\ref{rq:1.2}}     & Identify programming languages utilised.\\
        \textbf{\ref{rq:1.3}}     & Identify the application domain and example scenarios.\\
        \textbf{\ref{rq:2}}       & Identify stochastic, uncertain system application.  \\
        \textbf{\ref{rq:3}}       & Identify the research type facet.\\
        \bottomrule
    \end{tabularx}
\end{table}

\section{Study Results}
\label{sec:study-results}
The answers to the research questions are detailed in the following, with additional information presented first.
\subsection{\ref{si:1}: What are the demographic details of the research in \cbd in terms of citations, publication year, venue?}
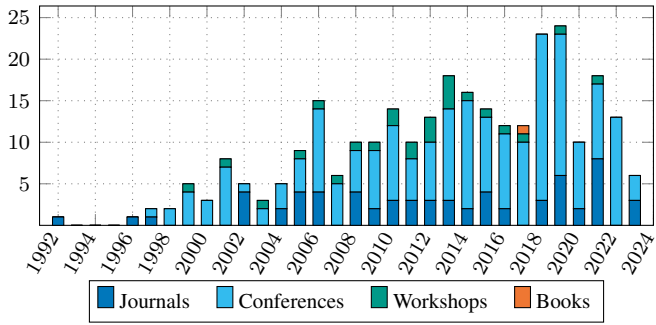
\begin{figure}
\centering
\begin{tikzpicture}
\pgfplotsset{grid style={dashed,gray}}
\pgfplotsset{minor grid style={dashed,gray}}
\pgfplotsset{major grid style={dotted,gray}}
  \begin{axis}[
    width=1.1\columnwidth,
    height=4.5cm,
    ybar stacked, ymin=0,  
    bar width=1.5mm,
    xtick=data,
    ytick={5, 10, 15, 20, 25, 30},
    xtick={1990, 1992,..., 2024},
    xmax=2024,
    xmin=1991,
    xtick distance=1,
    grid=major,
    yticklabel style={font=\footnotesize},
    xticklabel style={/pgf/number format/1000 sep=,rotate=60,anchor=east,font=\footnotesize},
    legend style={at={(0.5,-.25)},anchor=north, legend columns=-1, font=\footnotesize},
  ]
  \addplot [fill=pt_blue] table[x index=0,y index=2,col sep=tab] {data/rq1.dat};
  \addplot [fill=pt_cyan] table[x index=0,y index=3,col sep=tab] {data/rq1.dat};
  \addplot [fill=pt_teal] table[x index=0,y index=4,col sep=tab] {data/rq1.dat};
  \addplot [fill=pt_orange] table[x index=0,y index=5,col sep=tab] {data/rq1.dat};
  \legend{Journals\hspace*{1em},Conferences\hspace*{1em},Workshops\hspace*{1em}, Books}
  \end{axis}
  \end{tikzpicture}
    \caption{Publication trends over the years by venue}
    \label{fig:trend}
\end{figure}

Referring to \cref{fig:trend}, a clear trend emerges, showing a consistent rise in studies since 1997. Notably, while the pioneering paper by Meyer \cite{meyer:computer:92}, which introduced Design by Contract in the realm of reliable software, was published in 1992, the shift towards \cbd gained momentum only in the subsequent years. This increasing trend underscores this methodology's escalating relevance and widespread adoption across diverse fields like automotive and aerospace within the domain of dependable systems. Regarding publication venues, our analysis reveals a broad representation across various platforms, with conferences being particularly predominant.
Interestingly, the earliest primary study in our dataset also holds the distinction of being the most cited. This prominence is attributable to its foundational contribution to establishing a comprehensive set of principles for developing correct and reliable software. The methodologies proposed in this primary study laid the groundwork for subsequent research, fostering the development of new approaches. Other top-cited primary studies primarily owe their status to their longevity, as no discernible patterns emerge beyond the factor of time. For an in-depth examination of this seminal primary study and its influences, refer to the cited works in \cref{tab:citations}.
\begin{table}
    \centering
    \caption{Top 11 paper with the most citations}
    \label{tab:citations}
    \begin{tabularx}{\columnwidth}{@{}R{.6cm}XR{.3cm}L{1.1cm}@{}}
    \toprule
    \textbf{Paper}  &  \textbf{Title} & \textbf{Year} & \textbf{Citations} \\
    \toprule

\cite{meyer:computer:92} & Applying ``Design by Contract'' & 1992 & 3534 \\
\cite{jezequel:97} & Design by Contract: The Lessons of Ariane & 1997 & 414 \\
\cite{sangiovanni:12} & Taming Dr. Frankenstein: Contract-Based Design for Cyber-Physical Systems & 2012 & 403  \\
\cite{benveniste:07} & Multiple Viewpoint Contract-Based Specification and Design & 2007 & 260  \\
\cite{giannakopoulou:02} & Assumption generation for software component verification & 2002 & 238 \\
\cite{bobaru:08} & Automated Assume-Guarantee Reasoning by Abstraction Refinement & 2002 & 205 \\
\cite{bocchi:10} & A Theory of Design-by-Contract for Distributed Multiparty Interactions & 2010 & 185  \\
\cite{karaorman:99} & jContractor: A Reflective Java Library to Support Design by Contract & 1999 & 176  \\
\cite{feng:07} & On the Relationship Between Concurrent Separation Logic and Assume-Guarantee Reasoning & 2007 & 174 \\
\cite{kwiatkowska:10} & Assume-Guarantee Verification for Probabilistic Systems & 2013 & 170  \\
\cite{nuzzo:14} & A Contract-Based Methodology for Aircraft Electric Power System Design & 2010 & 170\\
    \bottomrule
    \end{tabularx}
\end{table}
\subsection{\ref{rq:1}: What are the design details of \gls{cbd}?}
\pgfplotstableread[row sep=\\,col sep=&]{
    interval          & carT & carD & carR \\
    Manufacturing     & 6  &  & 28.6 \\
    Power Systems     & 5  &  & 28.6 \\
    Communication     & 3  &  & 27.0 \\
    Robotics          & 38 & 38 & 22.2 \\
    Others            & 165  & 165 & 13.4 \\
}\mydataa

\pgfplotstableread[row sep=\\,col sep=&]{
    interval        & carT & carD & carR \\
    Java            & 64  & 0.1  & 0.2  \\
    Eiffel          & 5 & 3.8  & 4.9  \\
    C/C++           & 29 & 10.4 & 13.4 \\
    Python          & 11 & 17.3 & 22.2 \\
    Matlab          & 29  & 21.1 & 27.0 \\
    Others          & 10  & 22.3 & 28.6 \\
}\mydatab

\pgfplotstableread[row sep=\\,col sep=&]{
    interval        & carT & carD & carR \\
    Behavioural     & 98  & 0.1  & 0.2  \\
    Temporal        & 11 & 3.8  & 4.9  \\
    Safety          & 15 & 10.4 & 13.4 \\
    Stochastic      & 20  & 22.3 & 28.6 \\
}\mydatac

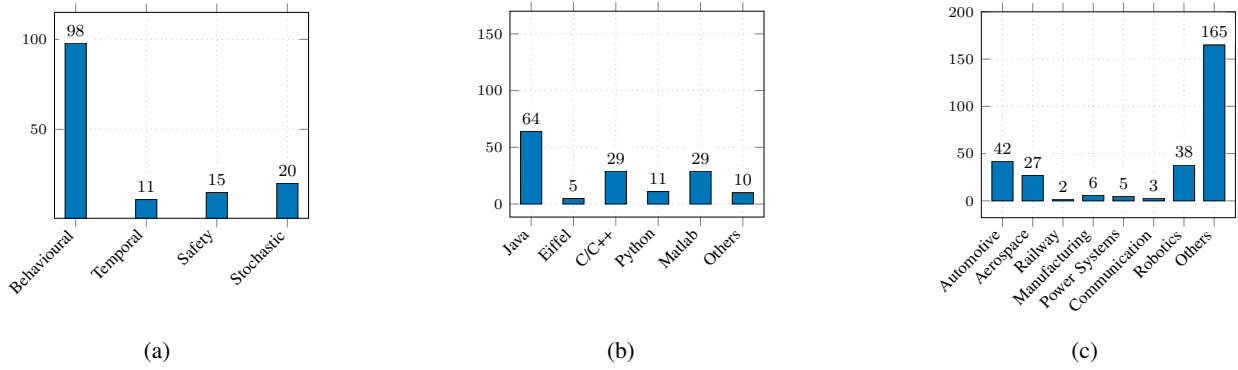
\begin{figure*}[h]
    \centering
    \begin{subfigure}[t]{0.32\textwidth}
        \centering
        \scalebox{0.8}{%
        \begin{tikzpicture}
            \begin{axis}[
                ybar,
                symbolic x coords={Behavioural,Temporal,Safety,Stochastic,None},
                xtick=data,
                ymax = 115,
                grid, grid style=dotted,
                nodes near coords,
                nodes near coords align={vertical},
                nodes near coords style={font=\footnotesize, text=black, fill=white}, 
                xticklabel style={rotate=45,anchor=east,align=center, font=\footnotesize,},
                width=\linewidth,
                ]
                \addplot [fill=pt_blue] table[x=interval,y=carT]{\mydatac};
            \end{axis}  
            \node[]() at (0, -15.1mm){\vphantom{X}};
        \end{tikzpicture}
    }
        \caption{}
        \label{fig:typeofcbd}
    \end{subfigure}
    \hfill
    \begin{subfigure}[t]{0.32\textwidth}
        \centering
        \scalebox{0.8}{%
        \begin{tikzpicture}
            \begin{axis}[
                ybar,
                symbolic x coords={Java,Eiffel,C/C++,Python,Matlab,Others,None},
                xtick=data,
                ymax=170,
                grid, grid style=dotted,
                nodes near coords,
                nodes near coords align={vertical},
                nodes near coords style={font=\footnotesize, text=black, fill=white}, 
                xticklabel style={rotate=45,anchor=east,align=center,font=\footnotesize,},
                width=\linewidth,
                ]
                \addplot [fill=pt_blue] table[x=interval,y=carT]{\mydatab};
            \end{axis}
            \node[]() at (0, -15.3mm){\vphantom{X}};
        \end{tikzpicture}
    }
        \caption{}
        \label{fig:lang}
    \end{subfigure}
    \hfill
    \begin{subfigure}[t]{0.32\textwidth}
        \centering
        \scalebox{0.8}{%
        \begin{tikzpicture}
            \begin{axis}[
                ybar,
                symbolic x coords={Automotive,Aerospace,Railway,Manufacturing,Power Systems,Communication,Robotics,Others},
                xtick= {Aerospace,Automotive,Railway,Manufacturing,Power Systems,Communication,Robotics, Others},
                grid, grid style=dotted,
                nodes near coords,
                nodes near coords align={vertical},
                nodes near coords style={font=\footnotesize, text=black, fill=white}, 
                ybar=-0.35cm,
                ymax=200,
                xticklabel style={rotate=45,anchor=east,align=center,font=\footnotesize,},
                width=1\linewidth,
                ]
                \addplot[fill=pt_blue]  coordinates { (Automotive,42) (Aerospace,27) (Railway,2)};
                \addplot [fill=pt_blue]table[x=interval, y=carT]{\mydataa};

            \end{axis}
        \end{tikzpicture}
    }
        \caption{}
        \label{fig:industry}
    \end{subfigure}
    \caption{\ref{rq:1}: \subref{fig:typeofcbd} Different contract types, \subref{fig:lang} used programming languages, \subref{fig:industry} the industry branches that apply \gls{cbd}.}
    \label{fig:three_graphs}
\end{figure*}
This section delineates the design specifics of \gls{cbd}.
\subsubsection{\ref{rq:1.1}: What varieties of \gls{cbd} are prevalent in the literature?}
This section elaborates on the diverse forms of \gls{cbd} observed in scholarly works. A notable challenge was the need for a standardised classification framework within the existing literature, manifested in diverse terminologies and types used across different studies. To mitigate this, we categorised contracts based on their core definitions as per \gls{cbd} guidelines. Here are the identified categories, along with their respective descriptions:
\begin{itemize*}
    \item \textbf{Behavioural Contracts:} These contracts outline the overarching behaviour of a system.
    \item \textbf{Temporal Contracts:} These specify the system's timing-related characteristics. 
    \item \textbf{Safety Contracts:} Focused on defining safe and unsafe states, rather than detailing system behaviour, they are crucial for preventing undesirable system conduct.
    \item \textbf{Stochastic Contracts:} Tailored for systems with uncertain or stochastic behaviours.
\end{itemize*}

Except for the primary studies shown in \cref{fig:typeofcbd}, the remaining papers are not included because their main focus is comparison methodologies within \gls{cbd} or providing tool support rather than directly proposing \gls{cbd}.
\subsubsection{\ref{rq:1.2}: Which programming languages are predominantly utilised for \cbd?}
This RQ aims to pinpoint the programming languages predominantly employed in \gls{cbd}. It was difficult to find clear references to programming languages in certain studies. As a result, we excluded such instances where pseudo-code or modelling languages like UML were utilised instead. Details can be seen in \cref{fig:lang}.
\subsubsection{\ref{rq:1.3}: What proportion of the primary studies employs \gls{cbd} in which industrial sector?}
The primary challenge was to extract specific industry contexts from the studies, especially those emphasising validation for algorithms over industry-specific applications. We categorised such instances under \emph{others}. Studies with ambiguous application areas, such as those concerning DC motor control algorithms, as in \cite{DerlerLTT13}, were classified under \emph{robotics}. We also found different industries related to dependable systems, such as \emph{automotive}, \emph{aerospace}, \emph{railway}, \emph{manufacturing}, \emph{power systems}, and \emph{communication}.
Notably, \emph{aerospace} and \emph{automotive} industries exhibit a higher prevalence of \gls{cbd} applications, as compared to sectors like, for instance, \emph{railway}, as detailed in \cref{fig:industry}.
\subsection{\ref{rq:2}: What design solutions have been proposed for \gls{dl}, \gls{rl}, \gls{rm}, and mitigation strategies?}
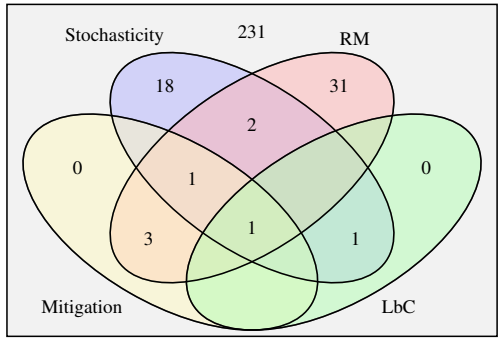
\begin{figure}
    \centering

    \tikzset{ellip/.style={ellipse, draw, thick,minimum height=3cm, minimum width=6cm, rotate=#1}}
    
\newcommand{\VennFour}[2][]{%
\resizebox{.74\columnwidth}{!}{%
\begin{tikzpicture}[#1]
    \draw[thick, fill=gray!10] (-4.5,-3.1) rectangle (4.5,3);
    
    \node[ellip=-35, label=160:Stochasticity, fill=blue!30, opacity=0.5] (A) at (0,0){};
    \node[ellip=35, label=20:RM, fill=red!30, opacity=0.5] (B) at (0,0){};
    \node[ellip=-30, label=270:Mitigation, fill=yellow!30, opacity=0.5] (C) at (-1.5,-1){};
    \node[ellip=30, label=270:LbC, fill=green!30, opacity=0.5] (D) at (1.5,-1){};
    \node[ellip=-35, draw=black] (A) at (0,0){};
    \node[ellip=35, draw=black] (B) at (0,0){};
    \node[ellip=-30, draw=black] (C) at (-1.5,-1){};
    \node[ellip=30, draw=black] (D) at (1.5,-1){};

    \coordinate(a)at(-1.6,1.5); \coordinate(b)at(1.6,1.5); \coordinate(c)at(-3.2,0); \coordinate(d)at(3.2,0);
    \coordinate(ab)at(0,.8); \coordinate(cd)at(0,-2.4); \coordinate(ac)at(-2.1,.55);
    \coordinate(bd)at(2.1,.55); \coordinate(ad)at(1.9,-1.3); \coordinate(bc)at(-1.9,-1.3);
    \coordinate(abc)at(-1.1,-.2); \coordinate(abd)at(1.1,-.2); \coordinate(acd)at(.75,-1.7); \coordinate(bcd)at(-.75,-1.7);
    \coordinate(abcd)at(0,-1.1); \coordinate(o)at(0,2.5);
    \foreach \s/\n in {#2} {\node at (\s){\n};}
\end{tikzpicture}}}
\VennFour{a/18, b/31, c/0, ab/2, bc/3, ad/1, abcd/1, abc/1, o/231, d/0}

    \caption{Distribution of proposed design solution}
    \label{fig:venn}
\end{figure}

Out of the analysed studies, only 23 addressed uncertain or stochastic system components, with just two studies (cf.~\cite{GuissoumaZS23,NaikN20}) proposing \gls{cbd} applications for \gls{rl} or \gls{dl}-based components, referred to as \gls{lbc}. 
Specifically, Naik and Nuzzo~\cite{NaikN20} propose a \gls{cbd}-based framework enabling scalable verification across system components to assess the robustness of systems incorporating neural networks, particularly in perception and control. 
Guissouma~et~al.~\cite{GuissoumaZS23} present a method for evaluating models against predefined safety contracts to ensure robustness and reliability, particularly in automotive applications such as Lane Keep Assist, for assessing the safety of machine learning models. Additionally, the paper proposes a mitigation strategy: when a safety contract is violated, the system automatically reverts to the previously validated model to maintain safe operation. This is the only study addressing all three aspects (cf.~\cref{fig:venn}). In the context of \gls{rm}, 38 studies proposed solutions for system monitoring during operation. Mitigation strategies for unexpected behaviours were discussed in only five studies; two examples include Sievers and Madni~\cite{Sievers2014}, who propose \emph{flexible contracts} using Hidden Markov Models to adapt to unexpected system behaviours, allowing the system to reconfigure itself. Frenkel~et~al.~\cite{frenkel2020} combine assume/guarantee reasoning with a repair mechanism, ensuring the program is automatically repaired to meet the desired specifications if verification fails.

\begin{figure}
    \centering
    \resizebox{\columnwidth}{!}{
    \begin{tikzpicture}
    \tikzset{help lines/.style={dashed, color=gray!50}}
    \def\pHeight{0.5\columnwidth}
	\def\pWidth{\linewidth}
	\def\cat{8}
	\def\foo{5}

    \foreach \c in {0,1,...,\cat} {
        \draw[thin,help lines] (\c,0) -- (\c,\foo);
    }
    \foreach \c in {0,1,...,\foo} {
        \draw[thin,help lines] (0, \c) -- (\cat,\c);
    }
    
     \draw[black] (0, 0) rectangle (\cat,\foo);
    
    \foreach \l [count=\c from 1] in {  {Automotive}, {Aerospace}, {Railway}, {Manufacturing}, {Power Systems}, {Communication}, {Robotics}, {Others}} {
        \node [rotate=45,anchor=east,align=right, inner sep=0pt] at (\c, -0.1) {\l};
    }

    \tikzstyle{stuff_fill}=[rectangle,fill=white!20]
  
    \foreach \l [count=\c from 1] in {{Probablilistic}, {\gls{lbc}}, {\gls{rm}},
    {Mitig. Strategies}} {
        \node[rotate=45,stuff_fill, anchor=east, xshift=-4pt] at (-0.1, \c) {\l};
    }
  
    \pgfplotstableread{data/rq3_new.dat}\table
    \pgfplotstablegetrowsof{\table}
    \pgfmathsetmacro{\M}{\pgfplotsretval-1}
    \pgfplotstablegetcolsof{\table}
    \pgfmathsetmacro{\N}{\pgfplotsretval-1}
  
    \foreach \row in {0,...,5}{
        \foreach \col in {0,...,\N}{
            \pgfplotstablegetelem{\row}{[index]\col}\of\table
                \ifnum\col=0
                    \xdef\x{\pgfplotsretval}
                \fi
                \ifnum\col=1
                    \xdef\y{\pgfplotsretval}
                \fi
                \ifnum\col=2
                    \xdef\radius{\pgfplotsretval}
                \fi
        }
        \definecolor{mycolor}{RGB}{\pdfuniformdeviate 256,%
            \pdfuniformdeviate 256,%
            \pdfuniformdeviate 256}
        \fill[pt_teal,opacity=.8]
        (\x,\y)circle({sqrt(\radius/3.1415)*0.22});
        \node[text = white, font=\footnotesize] at (\x,\y) {\radius};
    }

        \foreach \row in {6,...,\M}{
        \foreach \col in {0,...,\N}{
            \pgfplotstablegetelem{\row}{[index]\col}\of\table
                \ifnum\col=0
                    \xdef\x{\pgfplotsretval}
                \fi
                \ifnum\col=1
                    \xdef\y{\pgfplotsretval}
                \fi
                \ifnum\col=2
                    \xdef\radius{\pgfplotsretval}
                \fi
        }
        \definecolor{mycolor}{RGB}{\pdfuniformdeviate 256,%
            \pdfuniformdeviate 256,%
            \pdfuniformdeviate 256}
        \fill[pt_blue,opacity=.8]
        (\x,\y)circle({sqrt(\radius/3.1415)*0.22});
        \node[text = white, font=\footnotesize] at (\x,\y) {\radius};
    }
\end{tikzpicture}}%
\caption{Areas for CbD design solutions and number of papers.}
\label{fig:solvsapplic}
\end{figure}
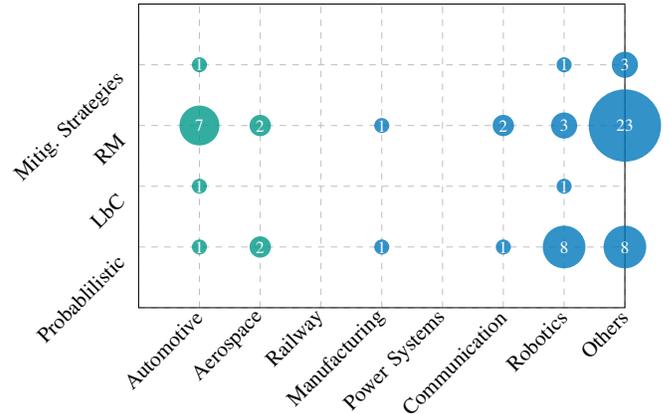

In addition, examining the industrial sectors where these proposed design solutions have been implemented reveals that solution approaches for \gls{rm} and probabilistic methods have been applied across various industries. We use probabilistic tagging here for \gls{cbd} in stochastic systems, but this does not directly relate to \gls{lbc} applications. We have two tags -- \gls{lbc} and probabilistic -- that together provide design solutions for stochasticity. Similarly, two primary studies proposing solutions for \gls{lbc} have been explicitly tested within the automotive and robotics sectors.    Further details are provided in \cref{fig:solvsapplic}.
\noindent%

An analysis of when these proposed solutions gained prominence within \gls{cbd} theory is also presented, with the temporal distribution of the primary studies shown in \cref{fig:designyears}. As illustrated in the figure, \gls{rm} solutions have been proposed within the \gls{cbd} framework for a considerable period. There are also primary studies dating back to 2010 that propose solutions for probabilistic components. In contrast, the situation is different for \gls{lbc}, as solutions for these components within \gls{cbd} only began to emerge in 2020.
\pgfplotstableread[row sep=\\,col sep=&]{
    interval        &RM  & Stoc & DLRL & Mit\\
    2001            &    1 & 0 & 0 & 0\\    
    2002            &    5 & 0 & 0 & 0\\             
    2003            &    0 & 0 & 0 & 0\\
    2004            &    0 & 0 & 0 & 0\\
    2005            &    2 & 0 & 0 & 0\\               
    2006            &    4 & 0 & 0 & 0\\               
    2007            &    0 & 0 & 0 & 0\\
    2008            &    2 & 0 & 0 & 0\\               
    2009            &    2 & 0 & 0 & 0\\               
    2010            &    1 & 2 & 0 & 0\\ 
    2011            &    3 & 0 & 0 & 1\\
    2012            &    0 & 2 & 0 & 0\\
    2013            &    3 & 2 & 0 & 0\\
    2014            &    3 & 1 & 0 & 1\\
    2015            &    2 & 1 & 0 & 1\\
    2016            &    0 & 2 & 0 & 0\\
    2017            &    0 & 1 & 0 & 0\\
    2018            &    2 & 0 & 0 & 0\\
    2019            &    3 & 6 & 0 & 0\\    
    2020            &    4 & 1 & 1 & 1\\
    2021            &    0 & 2 & 0 & 0\\
    2022            &    0 & 0 & 0 & 0\\
    2023            &    1 & 0 & 1 & 1\\
}\mydata
\pgfplotsset{height=12.9cm, width=9cm}
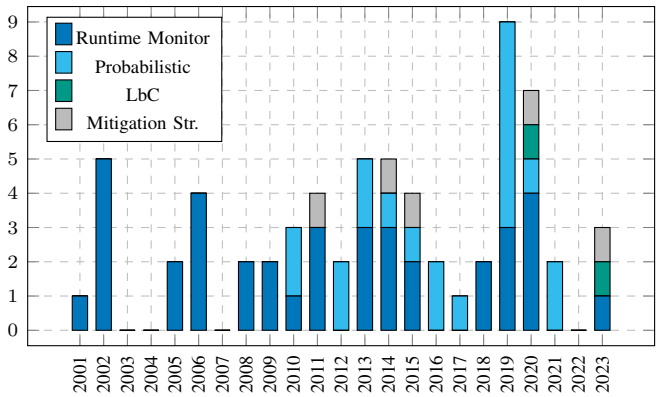
\begin{figure}
    \centering
    \resizebox{\columnwidth}{!}{
    \begin{tikzpicture}
        \begin{axis}[
            legend pos=north west,
            ybar stacked,
            bar width=.2cm,
            height=6.0cm,
            width=1.1\columnwidth,
            ytick={0,...,10},
            symbolic x coords = {2001,2002,2003,2004,2005,2006,2007,2008,2009,2010,2011,2012,2013,2014,2015,2016,2017,2018,2019,2020,2021,2022,2023},
            xtick = data, grid, grid style=dashed,
            xticklabel style = {rotate=90,anchor=east,align=center},
            enlarge y limits=0.05,
            legend style={nodes={scale=0.7, transform shape}}
        ]
            \addplot [fill=pt_blue] table[y=RM,x=interval]{\mydata};
            \addplot [fill=pt_cyan] table[y=Stoc,x=interval]{\mydata};
            \addplot [fill=pt_teal]table[y=DLRL,x=interval]{\mydata};
            \addplot [fill=pt_grey]table[y=Mit,x=interval]{\mydata};
            \legend{Runtime Monitor, Probabilistic, LbC, Mitigation Str.}
        \end{axis}
    \end{tikzpicture}}
    \caption{Studies containing design solutions over the years.}
    \label{fig:designyears}
\end{figure}
\subsection{\ref{rq:3}: What is the maturity level in different areas?}
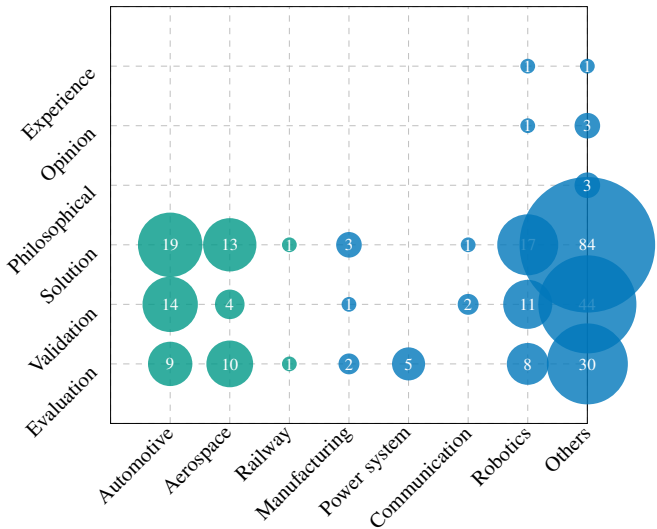
\begin{figure}
    \centering
    \resizebox{\columnwidth}{!}{
    \begin{tikzpicture}[]
    \tikzset{help lines/.style={dashed, color=gray!50}}
    \def\pHeight{0.5\columnwidth}
	\def\pWidth{\columnwidth}
	\def\cat{8}
	\def\foo{7}

    \foreach \c in {0,1,...,\cat} {
        \draw[thin,help lines] (\c,0) -- (\c,\foo);
    }
    \foreach \c in {0,1,...,\foo} {
        \draw[thin,help lines] (0, \c) -- (\cat,\c);
    }
    
     \draw[black] (0, 0) rectangle (\cat,\foo);

    \foreach \l [count=\c from 1] in {{Automotive}, {Aerospace}, {Railway}, {Manufacturing}, {Power system}, {Communication},{Robotics}, {Others}} {
        \node [rotate=45,anchor=east,align=right, inner sep=0pt] at (\c, -0.1) {\l};
    }

    \tikzstyle{stuff_fill}=[rectangle,fill=white!20]
  
    \foreach \l [count=\c from 1] in {{Evaluation}, {Validation}, {Solution}, {Philosophical}, {Opinion}, {Experience}} {
        \node[rotate=45, stuff_fill, anchor=east] at (-0.2, \c) {\l};
    }
  
    \pgfplotstableread{data/appl-area_new.dat}\table
    \pgfplotstablegetrowsof{\table}
    \pgfmathsetmacro{\M}{\pgfplotsretval-1}
    \pgfplotstablegetcolsof{\table}
    \pgfmathsetmacro{\N}{\pgfplotsretval-1}
  
    \foreach \row in {0,...,7}{
        \foreach \col in {0,...,\N}{
            \pgfplotstablegetelem{\row}{[index]\col}\of\table
                \ifnum\col=0
                    \xdef\x{\pgfplotsretval}
                \fi
                \ifnum\col=1
                    \xdef\y{\pgfplotsretval}
                \fi
                \ifnum\col=2
                    \xdef\radius{\pgfplotsretval}
                \fi
        }
        \definecolor{mycolor}{RGB}{\pdfuniformdeviate 256,%
            \pdfuniformdeviate 256,%
            \pdfuniformdeviate 256}
        \fill[pt_teal,opacity=.8]
        (\x,\y)circle({sqrt(\radius/3.1415)*0.22});
        \node[text = white, font=\footnotesize] at (\x,\y) {\radius};
    }
        \foreach \row in {8,...,\M}{
        \foreach \col in {0,...,\N}{
            \pgfplotstablegetelem{\row}{[index]\col}\of\table
                \ifnum\col=0
                    \xdef\x{\pgfplotsretval}
                \fi
                \ifnum\col=1
                    \xdef\y{\pgfplotsretval}
                \fi
                \ifnum\col=2
                    \xdef\radius{\pgfplotsretval}
                \fi
        }
        \definecolor{mycolor}{RGB}{\pdfuniformdeviate 256,%
            \pdfuniformdeviate 256,%
            \pdfuniformdeviate 256}
        \fill[pt_blue,opacity=.8]
        (\x,\y)circle({sqrt(\radius/3.1415)*0.22});
        \node[text = white, font=\footnotesize] at (\x,\y) {\radius};
    }
\end{tikzpicture}}%
\caption{Application areas and their maturity}
\label{fig:appl-area}
\end{figure}
An area is said to be mature if primary studies address diverse research type facets. To ensure a uniform classification, we employed the taxonomy developed by Wieringa~et~al.~\cite{Wieringa} that classifies studies into six facets:
\begin{itemize}
    \item \emph{Evaluation paper}: In this category, methods are implemented and assessed within substantial industrial, academic, or other practical environments.
    \item \emph{Validation paper}: These primary studies explore innovative techniques yet to be extensively implemented in industrial or academic settings.
    \item \emph{Solution proposal}: This class includes primary studies proposing a new solution or enhancing an existing problem-solving approach.
    \item \emph{Philosophical paper}: These studies offer fresh insights into current issues by developing a new taxonomy or conceptual framework or conducting literature reviews.
    \item \emph{Opinion paper}: Authors in this category present their viewpoints on contentious topics, often adopting a critical approach.
    \item \emph{Experience paper}: Primary studies in this class involve authors sharing their reflective evaluations based on their experiences in developing, implementing, and evaluating specific methods in engineering processes.
\end{itemize}
In our analysis, each selected primary study was categorised into one of these classes, and the application area of each primary study was identified. These findings are illustrated in \cref{fig:appl-area}.
An initial observation reveals a denser lower half of the graph, indicating a strong focus on solution-oriented studies. Over a third of the primary studies propose solutions or concepts that are yet to be implemented or validated. In contrast, the upper half is notably sparse, particularly in the section on personal experience studies. This limited coverage of real-world challenges and benefits in contract-based design highlights a significant gap in the literature.
\section{Discussion}
\label{sec:discussion}
Our research findings indicate many primary studies related to \gls{cbd} for various application areas. After conducting a voting process (as illustrated in \cref{fig:selection}), we were able to identify a total of 288 primary studies describing \gls{cbd} in terms of different aspects. In the following, we have summarised some of the findings and have extracted some key points from our systematic analysis of the selected studies.
\begin{itemize}
    \item \textbf{\gls{cbd}'s popularity has surged in the past two decades}: If we look at the findings in \cref{fig:trend}, we can see the increase in studies about this software paradigm between 1992 and 2019. After 2019, the number of studies decreased, but we can accept that this decrease is normal when considering the COVID-19 outbreak and its impact on the scientific world.
    \item \textbf{Java is the leading language for \gls{cbd} implementation:} 
        \begin{itemize}
            \item Although no mainstream programming language (except Eiffel) currently offers native support for \gls{cbd}, numerous plug-ins and tools have been developed to fill this gap. A prominent example is \gls{jml} \cite{leavens1998jml}, a Java extension that has significantly contributed to the widespread adoption of Java in the \gls{cbd} domain.
            \item C/C++ and Matlab have also been used extensively in \gls{cbd} implementations (cf.~\cref{fig:lang}).
            \item It is worth noting that many of the studies reviewed do not explicitly specify the programming language used.
        \end{itemize}
    \item \textbf{\gls{cbd} was most commonly used for system evaluation and monitoring:} 
    In many studies, contracts are used to define the behavioural, and temporal characteristics of the system, as well as to capture its uncertainty aspects. Additionally, A substantial portion of research also investigates \gls{cbd}’s role in system validation and monitoring, with 61 primary studies only on \emph{validation}.
    \item \textbf{Design solutions for \gls{lbc} within \gls{cbd} are rare:}
        \begin{itemize}
            \item While numerous primary studies focus on stochastic systems and \gls{rm}, the same is not true for \gls{lbc} despite their increasing relevance. Only two studies out of 288 (cf.~\cite{GuissoumaZS23,NaikN20}) have addressed such systems, both of which were conducted in recent years, as indicated in \cref{fig:venn,fig:designyears}.
            \item A limited number of studies provide mitigation strategies post-verification failure. These strategies vary, including contracts using Hidden Markov Models to adapt to system behaviours or repair mechanisms for systems using assume-guarantee reasoning.
            \item Only one primary study \cite{GuissoumaZS23} has presented \gls{cbd} for \gls{lbc}, \gls{rm}, and mitigation strategy.
            \item Finally, primary studies proposing solutions for \gls{rm} and stochastic systems are scattered across different industries. Primary studies with design solutions on \gls{lbc} have been implemented in robotics and automotive. For more detailed information, please see \cref{fig:solvsapplic}.
        \end{itemize}
    \item \textbf{Many primary studies lacked industry-specific testing of their designs:}
    Many primary studies tested their methods at the algorithmic level without specifying any industry sector, a trend particularly common in studies using \gls{cbd} for validation. Beyond this, the evidence indicates limited application of \gls{cbd} outside the automotive, aerospace, and manufacturing sectors, as shown in \cref{fig:industry}. Additionally, only one study within our research framework (cf.~\cite{tantivongsathaporn2006}) examined the impact of \gls{cbd} on the software development process.
    \item \textbf{Lack of primary studies critically reviewing \gls{cbd}:}
    We found that only a few studies provide subjective opinions on this topic, and no primary studies report practical experiences with \gls{cbd}, except in the fields of \emph{Robotics} and \emph{Others}. This scarcity of primary studies makes it challenging to evaluate the real-world application of this approach. In particular, in the area of \emph{Railways} (see \cref{fig:appl-area}), no work has been presented, aside from the \emph{Solution} and \emph{Evaluation} facets.
\end{itemize}
\subsection{Gaps and Future Research Directions}
According to our findings, this work presents gaps and future research directions in the following: 
\begin{itemize}
    \item A potential area for future research involves developing tools for various programming languages to facilitate \gls{cbd}. Currently, none of the well-known programming languages natively supports this methodology.
    \item To date, there has been no comprehensive study aimed at categorising contracts within \gls{cbd}. Future research could focus on conducting a detailed investigation into this area, ultimately providing a framework for \gls{cbd} categorisation.
    \item Research on \gls{cbd} that offers comprehensive solutions for learning-based components, runtime monitoring, and mitigation strategies is limited, particularly in the context of reliable systems. These systems, which often require continuous monitoring due to their stochastic nature, highlight the critical need for further research to elucidate this aspect of \gls{cbd}. Methodologies from existing studies can serve as foundational work, but further exploration is required to make these approaches more comprehensive.
    \item Another notable gap in the literature is the lack of studies on the effectiveness of \gls{cbd} across different industrial sectors. Further research is needed to assess the efficacy of \gls{cbd} implementations in these diverse fields.
\end{itemize}
\subsection{Threats to Validity} 
Like any empirical study, this research is subject to potential threats during data collection and processing, for which we have implemented several measures to mitigate risks.
\subsubsection{Threats to Construct Validity}
The initial design of our primary search query aimed to identify relevant studies on contracts. We adjusted the query for each database to maximise relevance. While the basic search terms performed well in the DBLP database, they produced many unrelated results in other databases. For example, the term \kw{contract} alone often led to irrelevant results from economics and social sciences. To better focus on safety-critical, dependable systems, we refined our search with specific keywords detailed in \cref{tab:search_strings}. This query underwent multiple revisions, especially for the IEEE Xplore and ACM Digital Library, to optimise the retrieval of pertinent literature.
\subsubsection{Threats to Internal Validity}
The refinement of our search strategy benefited significantly from the domain expertise of the third author. Keywords were meticulously selected to cover related concepts (\eg, \kw{NN}, \kw{neural network}) thoroughly. Despite these efforts, the effectiveness of our search in the IEEE Xplore database initially fell short. To enhance the quality of our findings, we made use of the advanced filtering options available in the database. These features allowed us to refine our results with greater precision. By carefully adjusting the filters, we were able to eliminate irrelevant data.
\subsubsection{Threats to External Validity}
A further threat concerns potential biases in selecting primary studies, particularly during voting. To minimise this risk, one independent author conducted data collection. Subsequently, the authors independently voted and selected studies, ensuring an unbiased selection process.
\section{Conclusion}
\label{sec:conclusion}
The increasing reliance on reliable systems in modern technology demands effective software design methodologies due to their complexity. \gls{cbd} has emerged as a flexible approach for both architectural and code-level development and validation.
Despite its recognition in academia, \gls{cbd} has seen limited adoption in the industrial sector, creating a noticeable gap. While many studies on \gls{cbd} theory exist in aerospace and automotive fields, there is a lack of experience and opinion studies. We aimed to assess the maturity of \gls{cbd}s industrial application through a systematic mapping study. From 1,221 primary studies across various databases, we selected 288 for detailed analysis.

Our study evaluated the maturity of \gls{cbd} in industrial applications, focusing on design solutions for learning-based components, runtime monitoring, and mitigation strategies in dependable systems. 

While \gls{cbd} has gained attention in academic literature over the past two decades, more primary studies on real-world implementations are needed. Most applications are concentrated in robotics, with limited representation in railways. Java is the main programming language used, though C/C++ and MATLAB are also noted, and many mainstream languages still lack native support for \gls{cbd}.

This study identified key research gaps: limited industry adoption of Component-based Development (CbD), insufficient programming language support, and a lack of comprehensive solutions for Lifecycle-based Computing (LbC). Additionally, categorizing types of \gls{cbd} was challenging due to a scarcity of prior research.
This research provides scholars with a foundational knowledge base on \gls{cbd}, highlighting insights and gaps while suggesting new directions for advancing reliable systems.
In our future work, we plan to investigate the reasons behind the limited industry adoption of \gls{cbd}. Our research will incorporate both scientific and experimental reports to gain a better understanding of this issue.
\begin{table}[]
    \centering
    \caption{Search Strings}
    \label{tab:search_strings}
    \resizebox{\columnwidth}{!}{%
    \begin{tabularx}{\columnwidth}{@{}X@{}}
        \toprule
        \textbf{Digital Library} and \textbf{Search String}\\
        \midrule
        \emph{\textbf{ACM Digital Library:}}
        ("Contract Based Design" OR "Design by Contract" OR "Assume Guarantee") AND ("Safety" OR "Behavior" OR "Behaviour" OR "Probabilistic" OR "Stochastic" OR "Timing" OR "Neural Network" OR "NN" OR "Artificial Intelligence" OR "AI" OR "Machine Learning" OR "ML" OR "Deep Neural Network" OR "DNN" OR "Recurrent Neural Network" OR "RNN" OR "Deep Learning" OR "DL") AND ("Automotive" OR "Autonomous Vehicle" OR "Autonomous" OR "Robotics" OR "Robot" OR "Mobile robot" OR "Cyber-Physical Systems" OR "Safety-Critical" OR "Embedded Systems" OR "Aviation" OR "Avionics" OR "Railway" OR "Manufacturing")\\
        \midrule
        \emph{\textbf{IEEE Xplore:}}
        ("Full Text Only": "Contract Based Design" OR "Full Text Only": "Design by Contract" OR "Full Text Only": "Assume Guarantee") AND ("Full Text Only": "Safety" OR "Full Text Only": "Behavior" OR "Full Text Only": "Behaviour" OR "Full Text Only": "Probabilistic" OR "Full Text Only": "Stochastic" OR "Full Text Only": "Timing" OR "Full Text Only": "Neural Network" OR "Full Text Only": "NN" OR "Full Text Only": "Artificial Intelligence" OR "Full Text Only": "AI" OR "Full Text Only": "Machine Learning" OR "Full Text Only": "ML" OR "Full Text Only": "Deep Neural Network" OR "Full Text Only": "DNN" OR "Full Text Only": "Recurrent Neural Network" OR "Full Text Only": "RNN" OR "Full Text Only": "Deep Learning", "Full Text Only": "DL" ) AND ("Full Text Only": "Automotive" OR "Full Text Only": "Autonomous Vehicle" OR "Full Text Only": "Autonomous" OR "Full Text Only": "Robotics" OR "Full Text Only": "Robot" OR "Full Text Only": "Mobile robot" OR "Full Text Only": "Cyber-Physical System" OR "Full Text Only": "Safety-Critical" OR "Full Text Only": "Embedded System" OR "Full Text Only": "Aviation" OR "Full Text Only": "Avionics" OR "Full Text Only": "Railway" OR "Full Text Only": "Manufacturing")\\  \textbf{Filter:} "formal verification", "formal specification", "program verification", "embedded systems", "safety-critical software", "object-oriented programming", "software architecture", "temporal logic", "cyber-physical systems", "contracts", "control engineering computing", "control system synthesis", "mobile robots", "software engineering", "specification languages", "aerospace computing", "production engineering computing", "Unified Modeling Language", "inference mechanisms", "learning (artificial intelligence)", "probability"\\
        \midrule
        \emph{\textbf{DBLP:}}
        "contract based design", "design by contract","assume guarantee"\\
        \bottomrule
    \end{tabularx}}
\end{table}
\bibliographystyle{IEEEtranS}
\balance
\bibliography{bibliography}

\begin{thebibliography}{10}
\providecommand{\url}[1]{#1}
\csname url@samestyle\endcsname
\providecommand{\newblock}{\relax}
\providecommand{\bibinfo}[2]{#2}
\providecommand{\BIBentrySTDinterwordspacing}{\spaceskip=0pt\relax}
\providecommand{\BIBentryALTinterwordstretchfactor}{4}
\providecommand{\BIBentryALTinterwordspacing}{\spaceskip=\fontdimen2\font plus
\BIBentryALTinterwordstretchfactor\fontdimen3\font minus
  \fontdimen4\font\relax}
\providecommand{\BIBforeignlanguage}[2]{{%
\expandafter\ifx\csname l@#1\endcsname\relax
\typeout{** WARNING: IEEEtranS.bst: No hyphenation pattern has been}%
\typeout{** loaded for the language `#1'. Using the pattern for}%
\typeout{** the default language instead.}%
\else
\language=\csname l@#1\endcsname
\fi
#2}}
\providecommand{\BIBdecl}{\relax}
\BIBdecl

\bibitem{baudry2001}
B.~Baudry, Y.~Le~Traon, and J.-M. J{\'e}z{\'e}quel, ``Robustness and
  diagnosability of oo systems designed by contracts,'' in \emph{Proceedings
  Seventh International Software Metrics Symposium}.\hskip 1em plus 0.5em minus
  0.4em\relax IEEE, 2001, pp. 272--284.

\bibitem{benveniste:07}
A.~Benveniste, B.~Caillaud, A.~Ferrari, L.~Mangeruca, R.~Passerone, and
  C.~Sofronis, ``Multiple viewpoint contract-based specification and design,''
  in \emph{Formal Methods for Components and Objects}.\hskip 1em plus 0.5em
  minus 0.4em\relax Springer Berlin Heidelberg, 2007, pp. 200--225.

\bibitem{benveniste:etal:fteda18}
A.~Benveniste, B.~Caillaud, D.~Nickovic, R.~Passerone, J.~Raclet,
  P.~Reinkemeier, A.~L. Sangiovanni{-}Vincentelli, W.~Damm, T.~A. Henzinger,
  and K.~G. Larsen, ``Contracts for system design,'' \emph{Found. Trends
  Electron. Des. Autom.}, vol.~12, no. 2-3, pp. 124--400, 2018.

\bibitem{bocchi:10}
L.~Bocchi, K.~Honda, E.~Tuosto, and N.~Yoshida, ``A theory of
  design-by-contract for distributed multiparty interactions,'' in \emph{CONCUR
  2010 - Concurrency Theory}.\hskip 1em plus 0.5em minus 0.4em\relax Springer
  Berlin Heidelberg, 2010, pp. 162--176.

\bibitem{DerlerLTT13}
P.~Derler, E.~A. Lee, S.~Tripakis, and M.~T{\"{o}}rngren, ``Cyber-physical
  system design contracts,'' in \emph{{ACM/IEEE} 4th International Conference
  on Cyber-Physical Systems, {ICCPS}'13}.\hskip 1em plus 0.5em minus
  0.4em\relax {ACM}, 2013, pp. 109--118.

\bibitem{feng:07}
X.~Feng, R.~Ferreira, and Z.~Shao, ``On the relationship between concurrent
  separation logic and assume-guarantee reasoning,'' in \emph{Programming
  Languages and Systems, 16th European Symposium on Programming, {ESOP} 2007},
  ser. Lecture Notes in Computer Science, vol. 4421.\hskip 1em plus 0.5em minus
  0.4em\relax Springer, 2007, pp. 173--188.

\bibitem{frenkel2020}
H.~Frenkel, O.~Grumberg, C.~Pasareanu, and S.~Sheinvald, ``Assume, guarantee or
  repair,'' in \emph{Tools and Algorithms for the Construction and Analysis of
  Systems: 26th International Conference, TACAS 2020, Proceedings, Part I
  26}.\hskip 1em plus 0.5em minus 0.4em\relax Springer, 2020, pp. 211--227.

\bibitem{bobaru:08}
M.~Gheorghiu~Bobaru, C.~S. P{\u{a}}s{\u{a}}reanu, and D.~Giannakopoulou,
  ``Automated assume-guarantee reasoning by abstraction refinement,'' in
  \emph{Computer Aided Verification}.\hskip 1em plus 0.5em minus 0.4em\relax
  Berlin, Heidelberg: Springer Berlin Heidelberg, 2008, pp. 135--148.

\bibitem{giannakopoulou:02}
D.~Giannakopoulou, C.~S. Pasareanu, and H.~Barringer, ``Assumption generation
  for software component verification,'' in \emph{17th {IEEE} International
  Conference on Automated Software Engineering {(ASE)} 2002}.\hskip 1em plus
  0.5em minus 0.4em\relax {IEEE} Computer Society, 2002, pp. 3--12.

\bibitem{GuissoumaZS23}
H.~Guissouma, M.~Zink, and E.~Sax, ``Continuous safety assessment of updated
  supervised learning models in shadow mode,'' in \emph{20th International
  Conference on Software Architecture, {ICSA} 2023 - Companion}.\hskip 1em plus
  0.5em minus 0.4em\relax {IEEE}, 2023, pp. 301--308.

\bibitem{hoare:cacm:69}
C.~A.~R. Hoare, ``An axiomatic basis for computer programming,'' \emph{Commun.
  {ACM}}, vol.~12, no.~10, pp. 576--580, 1969.

\bibitem{jezequel:97}
J.-M. Jazequel and B.~Meyer, ``Design by contract: the lessons of ariane,''
  \emph{Computer}, vol.~30, no.~1, pp. 129--130, 1997.

\bibitem{karaorman:99}
M.~Karaorman, U.~H{\"o}lzle, and J.~Bruno, ``jcontractor: A reflective java
  library to support design by contract,'' in \emph{Meta-Level Architectures
  and Reflection}.\hskip 1em plus 0.5em minus 0.4em\relax Berlin, Heidelberg:
  Springer Berlin Heidelberg, 1999, pp. 175--196.

\bibitem{Kitchenham}
B.~A. Kitchenham and P.~Brereton, ``A systematic review of systematic review
  process research in software engineering,'' \emph{Inf. Softw. Technol.},
  vol.~55, no.~12, pp. 2049--2075, 2013.

\bibitem{DBLP:KugeleOBCTH17}
S.~Kugele, P.~Obergfell, M.~Broy, O.~Creighton, M.~Traub, and W.~Hopfensitz,
  ``On service-orientation for automotive software,'' in \emph{2017 {IEEE}
  International Conference on Software Architecture, {ICSA}}.\hskip 1em plus
  0.5em minus 0.4em\relax {IEEE} Computer Society, 2017, pp. 193--202.

\bibitem{kwiatkowska:10}
M.~Z. Kwiatkowska, G.~Norman, D.~Parker, and H.~Qu, ``Assume-guarantee
  verification for probabilistic systems,'' in \emph{Tools and Algorithms for
  the Construction and Analysis of Systems, 16th International Conference,
  {TACAS} 2010. Proceedings}, ser. Lecture Notes in Computer Science, vol.
  6015.\hskip 1em plus 0.5em minus 0.4em\relax Springer, 2010, pp. 23--37.

\bibitem{leavens1998jml}
G.~T. Leavens, A.~L. Baker, and C.~Ruby, ``Jml: a java modeling language,'' in
  \emph{Formal Underpinnings of Java Workshop (at OOPSLA’98)}.\hskip 1em plus
  0.5em minus 0.4em\relax Citeseer, 1998, pp. 404--420.

\bibitem{mandrioli:meyer:aoose:91}
D.~Mandrioli and B.~Meyer, ``Design by contract,'' \emph{Advances in
  Object-Oriented Software Engineering}, p.~1, 1991.

\bibitem{mchugh2012interrater}
M.~L. McHugh, ``Interrater reliability: the kappa statistic,'' \emph{Biochemia
  medica}, vol.~22, no.~3, pp. 276--282, 2012.

\bibitem{meyer:jss:88}
B.~Meyer, ``Eiffel: {A} language and environment for software engineering,''
  \emph{J. Syst. Softw.}, vol.~8, no.~3, pp. 199--246, 1988.

\bibitem{meyer:computer:92}
------, ``Applying "design by contract",'' \emph{Computer}, vol.~25, no.~10,
  pp. 40--51, 1992.

\bibitem{NaikN20}
N.~Naik and P.~Nuzzo, ``Robustness contracts for scalable verification of
  neural network-enabled cyber-physical systems,'' in \emph{18th {ACM/IEEE}
  International Conference on Formal Methods and Models for System Design,
  {MEMOCODE} 2020}.\hskip 1em plus 0.5em minus 0.4em\relax {IEEE}, 2020, pp.
  1--12.

\bibitem{nuzzo:14}
P.~Nuzzo, H.~Xu, N.~Ozay, J.~B. Finn, A.~L. Sangiovanni-Vincentelli, R.~M.
  Murray, A.~Donzé, and S.~A. Seshia, ``A contract-based methodology for
  aircraft electric power system design,'' \emph{IEEE Access}, vol.~2, pp.
  1--25, 2014.

\bibitem{DBLP:RodriguesPMACBV17}
D.~Rodrigues, R.~de~Melo~Pires, E.~A. Marconato, C.~Areias, J.~C. Cunha, K.~R.
  L. J.~C. Branco, and M.~Vieira, ``Service-oriented architectures for a
  flexible and safe use of unmanned aerial vehicles,'' \emph{{IEEE} Intell.
  Transp. Syst. Mag.}, vol.~9, no.~1, pp. 97--109, 2017.

\bibitem{sangiovanni:12}
A.~L. Sangiovanni{-}Vincentelli, W.~Damm, and R.~Passerone, ``{Taming Dr.
  Frankenstein: Contract-Based Design for Cyber-Physical Systems},'' \emph{Eur.
  J. Control}, vol.~18, no.~3, pp. 217--238, 2012.

\bibitem{Sievers2014}
M.~Sievers and A.~M. Madni, ``A flexible contracts approach to system
  resiliency,'' in \emph{2014 IEEE International Conference on Systems, Man,
  and Cybernetics (SMC)}, 2014, pp. 1002--1007.

\bibitem{tantivongsathaporn2006}
J.~Tantivongsathaporn and D.~Stearns, ``An experience with design by
  contract,'' in \emph{2006 13th Asia Pacific Software Engineering Conference
  (APSEC'06)}.\hskip 1em plus 0.5em minus 0.4em\relax IEEE, 2006, pp. 335--341.

\bibitem{Wieringa}
R.~J. Wieringa, N.~A.~M. Maiden, N.~R. Mead, and C.~Rolland, ``Requirements
  engineering paper classification and evaluation criteria: a proposal and a
  discussion,'' \emph{Requir. Eng.}, vol.~11, no.~1, pp. 102--107, 2006.

\end{thebibliography}
\end{document}